\documentclass[%
aps,prb,preprint,groupedaddress,showkeys,
 amsmath,amssymb,
 aps,
nofootinbib]{revtex4-1}
\usepackage[flushleft]{threeparttable}
\usepackage{subcaption}
\usepackage{color}
\usepackage{graphicx}
\usepackage{dcolumn}
\usepackage{bm}
\usepackage{physics}
\usepackage{blindtext}
\usepackage{natbib}
\usepackage{float}
\usepackage{graphicx,wrapfig}

\begin{document}
\vspace*{0.35in}

\begin{flushleft}
{\Large
\textbf\newline{Simple analytic model for radiotherapeutic X-ray induced acoustic signal as a function of absorption parameters}
}
\newline
\\
Canberk Sanli\textsuperscript{1,*},
Esra Aytac Kipergil\textsuperscript{1},
Mehmet Burcin Unlu\textsuperscript{1,2},
\\
\bigskip
\bf{1} Department of Physics, Bogazici University,
34342 Bebek, Istanbul, Turkey
\\
\bf{2} Center for Life Sciences and Technologies, Bogazici University, 34342 Bebek, Istanbul, Turkey
\\
\bigskip
* canberk.sanli@boun.edu.tr

\end{flushleft}

\section*{Abstract}
Previous studies showed that it is possible make image reconstruction based on the dose dependence of the therapeutic XA (X-ray induced acoustic signal) amplitude which is then used to make dose mapping. We aimed to bring further explicit parametrization for the acoustic signal in terms of the absorption parameters, since this would mean encoding more information regarding the absorption process to XA signals. The first step is to obtain pressure waveform due to a point dose absorption by solving the thermo-acoustic equation governing the heat absorption-pressure induction process based on the analytic integration technique. Then, clinically relevant XA signal profile at the detection point is obtained by generalizing point-dose-gradient induced acoustic signal to surface-dose-gradient of a uniform spherical 3D dose distribution based on the reciprocity principle for pressure waves in fluid media. Therapeutic XA signal induced from the surface of the uniform spherical dose distribution due to X-ray irradiation onto $5x5$ $cm^2$ field of the water surface by $1~\mu s$ pulses delivering $1.7$ mGy/pulse is simulated in time and frequency domain. XA waves obtained in previous empirical studies are simulated and compared by means of shape and relative amplitude. Considering the previous studies on this subject, we believe that the significance of this study is the foundation of a novel and self-contained analytic approach to simulate the therapeutic X-ray acoustic waves based on the physical parametrization of the energy transfer process. This not only provides a better understanding of the physical phenomena underlying the medical technique in terms of the medically relevant parameters such as field size, pulse duration, absorbed dose per pulse etc. together with the physical assumptions used to obtain a solution to the photo-acoustic equation, but also brings consistent simulation results with previous experimental and k-Wave results.

\section{Introduction}

Pulsed X-ray beams produced in medical linacs (MeV range) or in X-ray tubes (keV range) induce ultrasonic signals~\cite{bowen91,screp} from which tomographic reconstruction of the irradiated region, called as X-ray induced Computed Tomography (XACT), is possible ~\cite{medphysxing,screp,hicklingfeasibility}. Hence, linac generated acoustic signals can be used for \textit{in vivo} dose monitoring during radiotherapy provided the higher spatial resolutions are achieved~\cite{medphysxing}, and the latter can be used to achieve deeper imaging than the current laser-based photoacoustic imaging techniques supply provided that low dose absorption does not pose a clinical risk~\cite{screp}. For both purposes, one needs to know the dependence of the acoustic wave profile on the absorption parameters. The linear dependence of pressure amplitude on the absorbed energy is already known from the general photoacoustic theory~\cite{biomedoptics} and qualitatively verified through XACT images \cite{hickling2016} by comparing XACT and ion chamber images. The dependence on the pulse duration is also qualitatively indicated by~\citet{screp}. 

Besides the plausibility of implementing XACT in clinical practice as X-ray imaging, X-ray induced acoustic (XA) waves can also be used as a real-time \textit{in vivo} dosimetry tool in radiotherapy. Since the X-ray sound solely depends on the X-ray absorption, any microscopic change in tissue properties due to radiotherapeutic beams can in principle be determined from the induced acoustic signals \cite{screp} if a complete description of the X-ray absorption regarding X-ray beam and tissue parameters is achieved. This would be an invaluable tool as it can increase the efficiency of radiotherapy treatments by monitoring the instantaneous treatment respond which may necessitate unplanned interventions. The feasibility of using XA waves as a dosimetry tool is emphasized in previous medical physics studies~\cite{medphysxing,hicklingfeasibility,screp}, and presciently by Askariyan~\cite{askariyan57} when he first mentioned the generation of acoustic signals due to the passage of ionizing radiation through matter. 

\citet{hickling2016} investigated this possibility by assuming a spatial profile for the heating function $H(\mathbf{r})$ which describes the instantaneous localized heating due to radiation exposure and defined through a `heat defect function' $k(\mathbf{r})$ in which all external absorption parameters such as pulse duration of the X-ray beam and absorption cross-section are implicit. Based on this formalism, they composed XACT images of which intensity profile formed by the relative pressure amplitudes of the ultrasound waves is used to obtain the dosimetric information which is then verified by separate measurements performed in an ion chamber. Consequently, the dose distribution is simulated with an accuracy depending on that in determining the heat defect, physical density, Gr\"{u}neisen coefficient, and properties of detection apparatus. 

This approach of using XACT images formed by the relative pressure amplitudes which are determined by the heating function defined above has two main theoretical deficiencies if its implementation for real-time \textit{in vivo} imaging in radiotherapy is considered. Firstly, since XACT images are based only on the relative pressure amplitudes, all the remaining information about the pressure signal is completely lost. This includes width and exact shape of the pressure profile. Secondly, the above definition for the heating function yields an absorption profile which does not give an explicit connection between X-ray beams dependent absorption parameters such as pulse duration and absorption cross section. Hence, one cannot relate the above-mentioned XACT images with these parameters which are important in dose absorption during radiotherapy~\cite{screp}.

These deficiencies do not cause any harm if XA waves are only to be used for imaging structural heterogeneities in matter such as lead blocks of different shapes in water \cite{hickling2014,hickling2016}, in chicken breast \cite{medphysxing}, or tiny piece of chicken bone in water phantom \cite{screp}. However, there exists no such contrast in shape and structure between normal and tumor tissue. Indeed, given that they are both treated as a water medium for most purposes, there exists no a priori reason why the amplitudes of the XA signals should significantly differ between tumor and normal tissue. Hence, XACT, as it is proposed, cannot be used to determine whether a detected XA signal is induced by the dose on a normal or tumor tissue which would be the main reason why one might perform a dose mapping during or after radiotherapy.

The aim of measuring induced acoustic signal is obtaining information concerning the absorption process, and the complete description of the X-ray absorption by deciphering the acoustic signal is possible, in principle. This means provided that accurate theoretical model with sufficient parameters is proposed, one can distinguish between XA signals due to different amount of dose absorption and different characteristics of medium where absorption takes place. Hence, an effort should be made to optimally model the dependence of pressure profile on the absorption process regarding the tissue and X-ray beam properties. Previous studies showed that it is possible make image reconstruction based on the dose dependence of pressure amplitude which is then used to make dose mapping. We aimed to bring further (than dose/amplitude dependence) explicit parametrization for acoustic signal in terms of the absorption parameters, since this would mean encoding more information regarding the absorption process to XA signals. For this purpose, we propose a novel analytic model for induced acoustic signals from the surface of a uniform spherical dose distribution due to radiotherapeutic X-ray exposure inside the tissue, and observed at an arbitrary position outside the distribution inside the same medium. The first step is to obtain pressure waveform due to a point dose absorption by solving the thermo-acoustic equation governing the heat absorption-pressure induction process based on the analytic integration technique. Then, clinically relevant XA signal profile at the detection point is obtained by generalizing point-dose-gradient induced acoustic signal to surface-dose-gradient of a uniform spherical 3D dose distribution based on the reciprocity principle for pressure waves in fluid media~\cite{landau1987fluid}. As a result, just like pressure amplitude mainly changes with the absorbed dose, we showed that pulse duration and absorption cross section also governs the waveform of the acoustic signal. It is an interesting curiosity whether these explicit dependencies of XA signal can be translated into the image reconstruction technique.

Thermo-acoustic model is first proposed to explain the induced acoustic signals due to energy transfer during the passage of high energy particles through matter; independently by~\citet{askariyan79} and~\citet{bowen77plovdiv}, and verified by~\citet{sulak78}. The solution to the thermo-acoustic equation is yielded by considering the deposited heat power per unit volume as a fictitious charge distribution and the observed pressure as `retarded potential'~\cite{askariyan79,sulak78,griffithsEM}. Then, by assuming a Gaussian distribution for instantaneous heat deposition, dependencies between pressure profile and particle cascade parameters are obtained~\cite{learnedgaussianheat} by direct integration. Only physical assumptions used to simplify the integration giving the point-source pressure wave solution are instantaneous heat deposition $\dot{T}˙ = T(r)\delta(t)$ due to X-ray exposure where $T(r,t)$ is the temperature rise at position $\vec{r}$ on the treatment plane, and sufficiently wide-band transducers so that absorption is frequency independent~\cite{askariyan79}. Validity of this technique in medical linac acoustic cases is already justified in medical physics literature~\cite{Baily}. Former physical assumption being true for any ionizing radiation can be elucidated by showing that the thermal relaxation time determined by the minimum dimensions of the area of heat evolution is essentially larger than the characteristic time of action of penetrating radiation determined by the pulse duration~\cite{lyamshev}, so that heat conduction can be neglected~\cite{wang2003heat}. As shown in the following section, this relation due to the implicit assumption of no sheer wave generation in the thermo-acoustic equation also serves as a bridge in expressing the absorption cross section in terms of a width parameter of a Gaussian curve which is a fit to the $5x5~cm^2$ square field on the treatment surface and forms the area of heat evolution. On the other hand, detection by immersion ultrasound transducer with a sufficiently large bandwidth so that frequency dependence of the absorption can be ignored is an priori assumption of our model. This is a reliable assumption considering the sharp peak of the therapeutic XA signal at a central frequency~\cite{hickling2014}. 

Although there have been many analytic simulations of the medical proton-acoustic emission after the confirmation of the thermo-acoustic model for treatment beams~\cite{Baily}, systematic study of dependency between physical absorption parameters and acoustic wave profile parameters is lacking~\cite{jones2014}. Analogously,~\citet{hickling2014} introduced a simulation workflow combining dose simulations and the induced acoustic wave transport simulation to form XACT images from which dosimetric information is extracted. This approach, however, does not reveal the explicit dependence between pressure profile and other important absorption parameters pulse duration and absorption cross section~\cite{screp}.

The method of solving thermo-acoustic equation is to assume a Gaussian distribution for the instantaneously deposited heat profile of which volume integral is then evaluated over the region of heat deposition at the retarded time, analogously to the derivation of analytic dependence of acoustic waves on absorption parameters from the thermo-acoustic model of high energy particles~\cite{learnedgaussianheat,askariyan79}. Analytic expression for the acoustic pressure profile generated by a point dose deposition inside the sample is then obtained as a function of the absorbed dose at that point, and absorption parameters pulse duration and absorption cross section. Dose gradient rate inducing acoustic pulses are determined by both beam energy and pulse repetition frequency, whereas pulse duration is solely determined by the linac operation~\cite{ppslinac73}. On the other hand, temperature rise at any point of the irradiated medium depends on the local absorption probability governed by the absorption cross section $\sigma_{abs}$ which is given by the product of mass energy attenuation coefficient with the atomic number density of the material. We approximate this temperature rise profile by a Gaussian curve of width parameter $\sigma$ which is fit to the square field size on the treatment surface and determines the minimum dimension of the area of heat evolution which is equal to the inverse of the mass energy attenuation coefficient for X-ray absorption~\cite{lyamshev}. Hence, given a uniform medium, absorption cross section for X-rays only depends on the field size on the surface which on the other hand defines the width parameter of Gaussianly distributed heat absorption profile in our model. In other words, given a uniform medium, absorption cross section only depends on the width parameter of Gaussianly distributed heat absorption profile in our model.

In a clinically applicable case the acoustic signal reaching the detector is a superposition of this point-wise generated waves in accordance with the 3D dose distribution due to X-ray exposure, and the detector is located at an arbitrary position outside this distribution. Hence, we generalized the analytic expression for point-dose-gradient induced acoustic signal during X-ray radiotherapy to surface-dose-gradient of a uniform spherical 3D dose distribution based on the reciprocity principle, and ignoring the attenuation of the acoustic signal due to small attenuation coefficient, exponential decay and close-range detection. This generalization enables the consistent comparison of the simulated acoustic waveforms based on our analytic model with the plausible clinical scenario where isodose curve is approximately spherical due to depth-dose curve for X-rays~\cite{spherical_dose} and induces acoustic signals as previously observed and simulated by~\citet{hickling2016}. Furthermore, dependence of the XA signal on the dose rate, absorption cross section, and pulse duration are quantitatively investigated by simulations in the therapeutic range of these parameters.

The approach taken here can be used for any radiation induced acoustic wave mechanism in general, given the expression for the initial pressure rise depending on the local energy absorption properties, particularly for proton induced acoustic waves of which the initial pressure rise is similarly given~\cite{ahmadxiangprot}.

\section*{Materials and Methods}

For a point source of heat which is located at $\vec{r}$ and described by a macroscopic heat deposition function $H(\vec{r},t)$ associated with the temperature rise $T(\vec{r},t)$ at any point $\vec{r}$ on the calculation plane which is caused by the dose gradient around a point $\eta_0$ taken as the origin of our coordinate system, the generation and propagation of XA wave are given by the photoacoustic equation~\cite{biomedoptics,tutoriol_PAI,morse}

\begin{eqnarray}
(\nabla^2 - \frac{1}{v_s^2}\frac{\partial^2}{\partial t^2})p(\vec{r},t)&=&-\frac{\beta}{\kappa v_s^2}\frac{\partial^2 T}{\partial t^2} \\
(\nabla^2 - \frac{1}{v_s^2}\frac{\partial^2}{\partial t^2})p(\vec{r},t)&=&-\frac{\beta\rho C_v}{C_p}\frac{\partial^2 T}{\partial t^2} 
\end{eqnarray}
of which solution $p(\vec{r},t)$, i.e. the pressure at any point $\vec{r}$ and at time $t$, is most simply expressed in terms of the volume integral over the region of absorption (heat deposition), i.e. space $V$ which is formed by a point source of heat located at $\vec{r'}$ and bounded by the stress confinement sphere $S_r^R = v_s \tau$ where $v_s$ is the speed of sound in the medium and $\tau$ is the pulse duration, at the retarded time $t-{\vert \vec{r}}-{ \vec{r'}\vert}/v_s$ \cite{sulak78,Baily,terunuma2007,kruger_reconstr}

\begin{equation}
p(\vec{r},t)=\frac{\beta \rho C_v}{4\pi C_p}\int_{V}\frac{dV'}{{\vert{\vec{r}}-{\vec{r'}}}\vert}\frac{\partial^2}{\partial t^2}T{\bigg{(}}{\vec{r'}},t-\frac{{\vert \vec{r}}-{ \vec{r'}\vert}}{v_s}{\bigg{)}}
\end{equation} 

which can be further reduced to the integration over the spherical surface of radius $R=v_s \tau$ with its center at $\vec{r}$ ~\cite{askariyan79}
\begin{equation} \label{pres_surface}
p(\vec{r},t)=\frac{\beta \rho C_v}{4\pi C_p}v_s^2\frac{\partial}{\partial R}\int_{S_r^R}dS \frac{T(\vec{r},t)}{R}
\end{equation}
by assuming instantaneous heat deposition $\dot{T}˙ = T(r)\delta(t)$ due to X-ray exposure, and sufficiently wide-band transducers so that absorption is frequency independent, where $\beta$ is thermal coefficient of volume expansion, $\rho$ is the mass density of soft tissue, and $C_p$ and $C_v$ denote the specific heat capacities at constant pressure and volume which together define the isothermal compressibility $\kappa=C_p/(\rho v_s^2 C_v)$.

\begin{figure}[!h]
\includegraphics[scale=1]{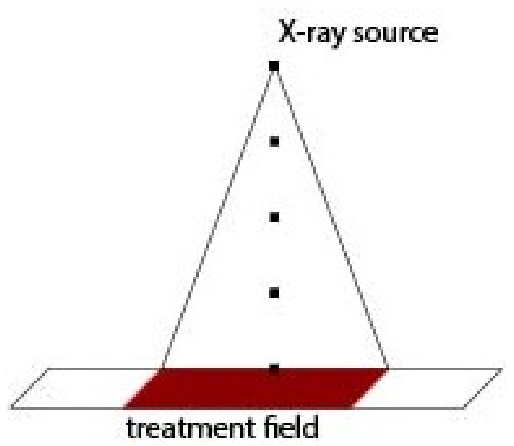}
\includegraphics[scale=0.7]{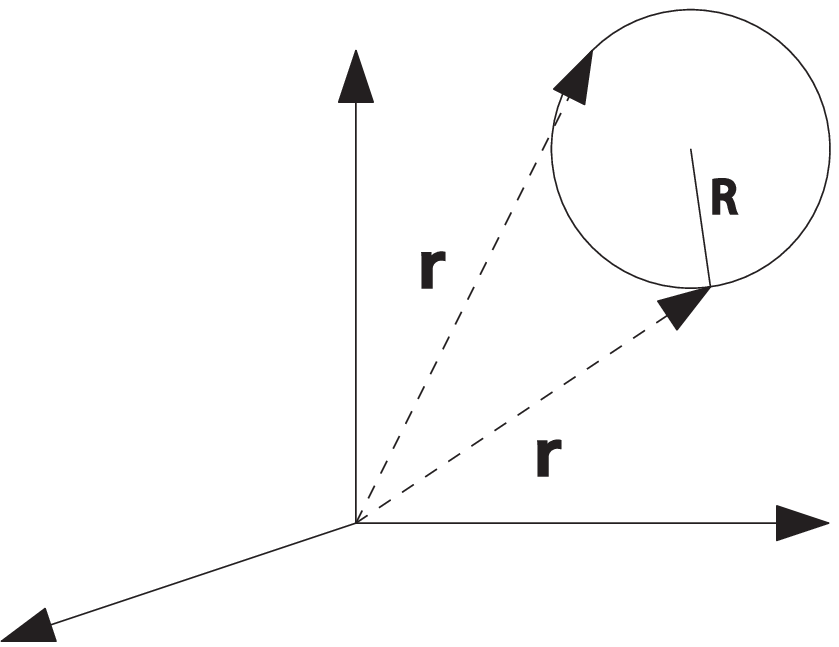}
\caption{\bf Treatment Field and Integration Region}
\label{integration}
\end{figure}

Here, the temperature rise $T(\vec{r},t)$ is related to the heating function $H(\vec{r},t)$ defined as the thermal energy converted per unit volume per unit time by \cite{biomedoptics}
\begin{equation} \label{Heating_and_Temp}
\rho C_v \frac{\partial T(\vec{r},t)}{\partial t}=H(\vec{r},t) 
\end{equation}
which is expressed in terms of a product of spatial and temporal Gaussian profiles as 
\begin{equation} \label{Heatprofile_eqn}
H(\vec{r},t)=\frac{p_0 C_p}{\beta v_s^2}e^{-r^2/2\sigma^2}\frac{1}{\sqrt{2\pi\tau^2}}e^{-t^2/2\tau^2} ,
\end{equation}
on the macroscopic treatment surface taken as the $5x5~cm^2$ square field, in order to reveal the dependence of pressure profile on the X-ray absorption parameters: pulse duration-$\tau$, and absorption cross section-$\sigma_{abs}$ which is defined by the width parameter $\sigma$ of the Gaussian fit to the treatment surface.

Integrating Eq (\ref{Heatprofile_eqn}) gives
\begin{equation}
\rho C_v T(\vec{r},t)=\frac{p_0 C_p}{\beta v_s^2}\frac{e^{-r^2/2\sigma^2}}{\sqrt{2\pi\tau^2}}\int_{0}^{t} e^{-t'^2/2\tau^2}dt'
\end{equation}
where
\begin{equation}
\int_{0}^{t} e^{-t'^2/2\tau^2}dt'=\tau\sqrt{\frac{\pi}{2}} \mathbf{erf}(t/\sqrt{2}\tau)
\end{equation}
and hence
\begin{equation}
\rho C_v T(\vec{r},t)=\frac{p_0 C_p}{\beta v_s^2}\frac{e^{-r^2/2\sigma^2}}{2}\mathbf{erf}(t/\sqrt{2}\tau) \label{Temp_eqn}
\end{equation}

Here, we note that $T(\vec{r},t)$ is not exactly constant on the integration surface $S_r^R$ centered at r, with radius $R=v_s\tau$, as in Fig~\ref{integration}. Since $R$ is defined by the region of stress confinement that is defined by the pulse duration, and hence is small whereas $r$ is always of the macroscopic size; $T(\vec{r},t)$ is assumed to be constant and moved outside the integration in Eq~(\ref{pres_surface}) so that after substituting (\ref{Temp_eqn}) into (\ref{pres_surface}) we get
\begin{eqnarray}
p(\vec{r},t)&=&\frac{p_0 e^{-r^2/2\sigma^2}}{8\pi}\mathbf{erf}(t/\sqrt{2}\tau) \frac{\partial}{\partial R}\int_{S_r^R} R^2 (1/R) sin\theta d\theta d\phi \\
p(\vec{r},t)&=&\frac{p_0}{2}e^{-r^2/2\sigma^2}\mathbf{erf}(t/\sqrt{2}\tau) \label{photoacoustic_pres} \end{eqnarray}

To sum up, the idea is to consider the macroscopic temperature rise distribution as a Gaussian curve lasting for a small time of a pulse duration $\tau$ defining at any point $\vec{r}$ a small region of stress confinement $S_r^R$ of radius $R=v_s \tau$ over whose surface $T(r,t)$ remains constant. Thus, Eq (\ref{photoacoustic_pres}) which can also be seen as the special case of the previously established photoacoustic deconvolution algorithm~\cite{deconvolution} with Gaussian deposition of heat~\cite{learnedgaussianheat} gives the pressure rise at point $r$ due to initial pressure rise $p_0$ at the origin. This defines a pressure profile (field) $p(r,t)$ due to a single point heat source over the macroscopic treatment region restricted by the field size $5x5~cm^2$. We neglect the attenuation of the induced ultrasound due to small attenuation coefficient, exponential decay and close-range detection to conclude that a transducer located nearby this region would observe the same profile $p(r,t)$. We will generalize this pressure profile induced by a point-dose-gradient to the waveform $p^{3D}(\vec{\eta},t)$ induced from a dose gradient on the surface of a uniform dose distribution as a function of the distance to the surface of the sphere from the transducer location $\vec{\eta}$.

For an X-ray induced acoustic wave (XA), the initial pressure rise is given by~\cite{hickling2014}
\begin{equation} \label{initial_pressure_rise}
p_0(\vec{r})=\Gamma \rho \tilde{D}(\vec{r})
\end{equation}
where $\tilde{D}$ is the absorbed dose per pulse at the position $\vec{r}$ on the treatment surface, and $\Gamma$ being the Gruneisen parameter describing conversion efficiency between deposited heat energy and pressure~\cite{hickling2016} is defined in terms of thermal coefficient of volume expansion $\alpha$ and specific heat capacity at constant pressure $C_p$ as $\Gamma = \frac{\beta}{C_p}v_s^2$ where $v_s$ is the speed of sound in the medium, and taken to be uniform; $\Gamma=0.15$, since detection is assumed to be realized in a homogeneous water tank at room temperature. Hence, acoustic pressure profile of XA wave at ($\vec{r}$,t) due to point heat source located at $\vec{r}_0$ is given by 
\begin{eqnarray}
p(\vec{r},t)&=&\frac{\Gamma \tilde{D}(\vec{r}_0) \rho}{2}~e^{-r^2/2\sigma^2}~\mathbf{erf}(t/\sqrt{2}\tau) \label{pres_perpulsedoseXA} 
\end{eqnarray}
where $\tilde{D}(\vec{r}_0)=\tilde{D}$ is the absorbed dose per pulse at point $\vec{r}_0$ on the treatment surface and is the first order approximation to the nearby dose in the region of interest, i.e. $D(\vec{r}) \simeq D(\vec{r}_0)$ for ${r} < 5~cm$, and determined by the medical range of dose rates and effective pulse repetition frequencies at which linacs are operated. To avoid any confusion concerning the range of the parameters, we again emphasize that $t$ defines a microscopic time related with the stress confinement whereas $r$ defines the macroscopic region characterized by a Gaussian of width $\sigma$ given by the field-size on the surface, and they together form the temperature profile in space and time of which surface integral at each point over the region of stress confinement as in Eq~(\ref{pres_surface}) forms the pressure profile on the treatment surface in the course of several pulse durations. 

In a realistic case however the acoustic signal reaching the detector would be the superposition of our point-wise generated waves given by Eq~(\ref{pres_perpulsedoseXA}) in accordance with the 3D dose distribution due to X-ray exposure, and the detector would be in an arbitrary position from the dose gradient that induces XA signal. To achieve this, we first assumed 3D spherical uniform dose distribution inside the sample in a region of $l_{min}^3~cm^3$ where $l_{min}$ is the minimum dimension of the area of heat evolution being consistent with the relative dose-depth curve for X-rays~\cite{spherical_dose,podgorsak2005} together with the assumption of isotropic media, and determined by the radius of the radiation spot $a/2$ for X-ray irradiation by $l_{min}\approx a/2 \approx \sigma$~\cite{lyamshev} where $a$ is the field size, as will be explained. Pressure waves are then induced due to dose gradient on the surface $S$ of this spherical region. Our aim is to calculate the pressure measured by a transducer located nearby the sphere. Let us put the transducer first at the center of the sphere. According to the reciprocity principle, pressure at this center due to all point sources (dose per pulse) located on the sphere is equal to the total pressure on the spherical surface due to point source (dose per pulse) located at the center. This is equivalent to assuming that point sources on the sphere is distributed homogeneously. Hence, to find the pressure at the center of the spherical dose distribution, $p^{3D}(\vec{r_0},t)$, we equivalently find the average pressure field due to a point source Eq. (12) on the surface $S$.

\begin{eqnarray}
p^{3D}(\vec{r_0},t)&=&\frac{1}{4 \pi (a/2)^2}\int_{S} \frac{\Gamma \tilde{D}(\vec{r_0}) \rho}{2}~e^{-\eta^2/2\sigma^2}~\mathbf{erf}(t/\sqrt{2}\tau)da \label{pres_3d_int}\\
p^{3D}(\vec{r_0},t)&=&\frac{\Gamma \tilde{D} \rho}{2}~\frac{\eta^2}{\sigma^2}~e^{-\eta^2/2\sigma^2}~\mathbf{erf}(t/\sqrt{2}\tau) \label{pres_3d}
\end{eqnarray}
where $\vec{\eta}$ is the distance to the surface of the sphere from the transducer location, and equal to the radius of the sphere for the case where the transducer is located at the center.

Second approximation is that ultrasound attenuation is negligible due to small attenuation coefficient, exponential decay and close-range detection. Hence, as we move the transducer radially outside the sphere, at a radial distance $\vec{\eta}$ from the surface, it will measure the signal given by

\begin{equation}
p^{3D}(\vec{\eta},t)=\frac{\Gamma \tilde{D} \rho}{2}~\frac{\eta^2}{\sigma^2}~e^{-\eta^2/2\sigma^2}~\mathbf{erf}(t/\sqrt{2}\tau) \label{pres_3d_sphere}
\end{equation}

Eq (\ref{pres_3d_int}) is a general expression for the pressure profile induced by the dose gradient at the surface $S$ of a 3D uniform dose distribution with a center at $r_0$ taken as the detection point. For a uniform spherical dose distribution, it yields Eq (\ref{pres_3d_sphere}) as the pressure profile measured by the transducer at a distance $\vec{\eta}$ from the surface of the sphere. We note that $\eta$ takes values between $d$ and $2l_{min}+d$ where $d$ is the closest distance to the sphere from the transducer and $2l_{min}=a$ is the field size and forms pressure profile at the transducer location as given below. This can be consistently compared with the previous X-ray induced acoustic studies, particularly with the study by \citet{hickling2016} where both experimental and k-wave simulated acoustic signals reaching from boundaries of various uniform 3D dose distributions in pure water due to different collimation of the X-ray field.

Dose gradient per pulse $\tilde{D}$ in Eq (\ref{pres_3d_sphere}) which is the main parameter of interest in X-ray acoustics can be determined via two approaches. One alternative is to use clinical dose values per fraction which are required in X-ray radiotherapy to be in between $1.7$ and $2.5$ Gy~\cite{typesaccelerator} and are verified through various experimental techniques~\cite{podgorsak2005, handbook_radiotherapy, phantom1, phantom2, ionc1, mosfet, ppslinac2016}, and also by Monte-Carlo simulations and analytical modeling~\cite{monte1,monte2,fastsimple,taddei_craniospinal}. This determines a dose gradient at the surface of the dose distribution due to the difference between the amount of dose in the treatment and the nearby regions. This difference is approximately 2 Gy fraction~\cite{fastsimple} and corresponds to $\tilde{D}=1.7$ mGy/pulse by taking the effective pulse repetition frequency as 60 Hz~\cite{ppslinac73,ppslinac2016} and macro-pulse duration to deliver 2 Gy fraction as 20 seconds~\cite{ppslinac2016}. Alternatively, one can use the pulse repetition frequency value directly with the dose rate taken as 6 Gy/min in the operational range at which medical linacs are operated in radiotherapy~\cite{typesaccelerator} to get the same dose-per-pulse value. Either way we deduce that radiotherapeutic dose gradient per pulse on the sphere can vary in the the interval 1 to 2 mGy/pulse as also taken in the previous therapeutic XA studies~\cite{medphysxing,hickling2014,hickling2016} and we simulate the acoustic signals accordingly.

Ultrasound-absorption dependence is investigated in terms of three parameters: absorbed dose per pulse, pulse duration and the absorption cross section. Range of pulse duration is taken in between 1-5 $\mu s$ as this is the standard range in which medical linacs are operated since their first appearance~\cite{ppslinac73,ppslinac2016}. Heat deposition function (and hence the temperature rise) function $H(\vec{r},t)$ is approximated as a Gaussian curve defined on all the calculation plane in Fig \ref{integration}, as given by Eq~\ref{Heatprofile_eqn}. The width parameter $\sigma$ of this Gaussian is calculated as 
\begin{equation} \label{eqn::sigma}
\sigma=\frac{a}{2\sqrt{2ln2}}
\end{equation}
by fitting a Gaussian curve centered at the center of the square and of which full width at half maximum intersects the edge of the square field of size 'a'. On the other hand, since absorption cross section is given by $\sigma_{abs}=\mu A /(\rho N_A)$, where $\rho$ is the mass density, $\mu$ is the X-ray absorption coefficient, and $N_A$ and $A$ are Avogadro number and atomic number, respectively, it is parametrized only by attenuation coefficient $\mu$, or equivalently by the minimum dimension of the area of heat evaluation $l_{min}$ given by $l_{min}=\mu^{-1}\approx a/2$ for X-rays~\cite{lyamshev}for a given medium, where $a/2$ is the radius of the radiation spot at the surface depending on the size of the treatment field which is taken as $5x5~cm^2$ in this study. The existence of this minimum dimension which is at the order of the field size (or $\sigma$) on the surface is due to the implicit assumption in the photo-acoustic equation that pulse duration is small enough to neglect the viscous effects so that we have $l_{min} >> 1/(\tau \nu)^{1/2}$, i.e. the minimum dimensions of the area of heat evolution (thermal source) must be essentially larger than the depth of wave penetration which is characterized by $1/(\tau \nu)^{1/2}$ where $\tau$ is pulse duration and $\nu$ is kinematic coefficient of viscosity. Therefore, in X-ray radiotherapy, assuming a uniform medium, absorption cross section is defined solely by the field size, and hence by the Gaussian width parameter $\sigma$ of the heating function. This means that variation in the $\sigma$ can be interpreted in two ways. That could be taken as a change in the field size, or a change in the medium, once one of them is fixed. In both cases, the effect can be observed in the pressure waveform and simulated by our simple analytic model. Corresponding pressure profiles as a function of dose gradient on the surface of a uniform spherical dose distribution per pulse, pulse duration, and absorption cross section as given by Eq~(\ref{pres_3d_sphere}) are plotted using MATLAB, Mathworks Inc. Based on these profiles, the central frequency $f_0$ of the signal is determined by applying Fast Fourier Transform (FFT) from which the noise level of a generic transducer with uniform efficiency $\eta(f)\approx\eta(f_0)=0.5$ and with size $A=3~cm^2$  to detect this signal is calculated by using ~\cite{noise_NEP,screp}
\begin{equation} \label{NEP_equation}
NEP=\sqrt{k_B T\Bigg[1+\frac{F_n}{\eta(f_0)}\Bigg]Z_a/A}~.~\sqrt{f_0}
\end{equation}
where $k_B$ is Boltzmann constant, $T$ is the absolute temperature, $F_n=2$ is the noise factor of the amplifier, and $Z_a$ is the characteristic acoustic impedance of the medium taken as water.

Similarly, one can use Eq~(\ref{pres_perpulsedoseXA}) in order to plot acoustic signals induced by proton beams used in proton radiotherapy as well, by directly implementing the dose values at the Bragg peak~\cite{koehlerprotondose,larssonprotondose,classieprotondose}, which is also analytically approximated by~\citet{Bortfeld97}.

\section*{Results}

Results contain two parts. First, we investigated the pressure signals induced by different amounts of dose gradients per pulse $\tilde{D}$ on the surface of a uniform spherical dose distribution. For this purpose, we fixed pulse duration $ \tau = 1 ~\mu s$  and the width parameter $\sigma = 2.12$ cm, and by plotting the Eq~(\ref{pres_3d_sphere}) for different values of dose per pulse values in the clinical range of 1 to 2 mGy/pulse with $\eta=-5:(15/2)10^{-4}:10$ and $t=-10^{-5}:10^{-9}:10^{-5}$, we obtained XA signal measured by a transducer at a distance $\eta$ from the surface of the sphere. Secondly, for a fixed amount of clinical dose gradient per pulse of 1.7 mGy/pulse on the surface, the dependencies of the pressure signal on $\tau$ and $\sigma$ are obtained by plotting Eq~(\ref{pres_3d_sphere}) for pulse duration values of $1-5~\mu s$ at a fixed $\sigma=2.12$ cm and for different width parameters 2.12, 1, 0.5 cm at a fixed $\tau=1~\mu s$.

\begin{figure}[!h]
\includegraphics[scale=0.55]{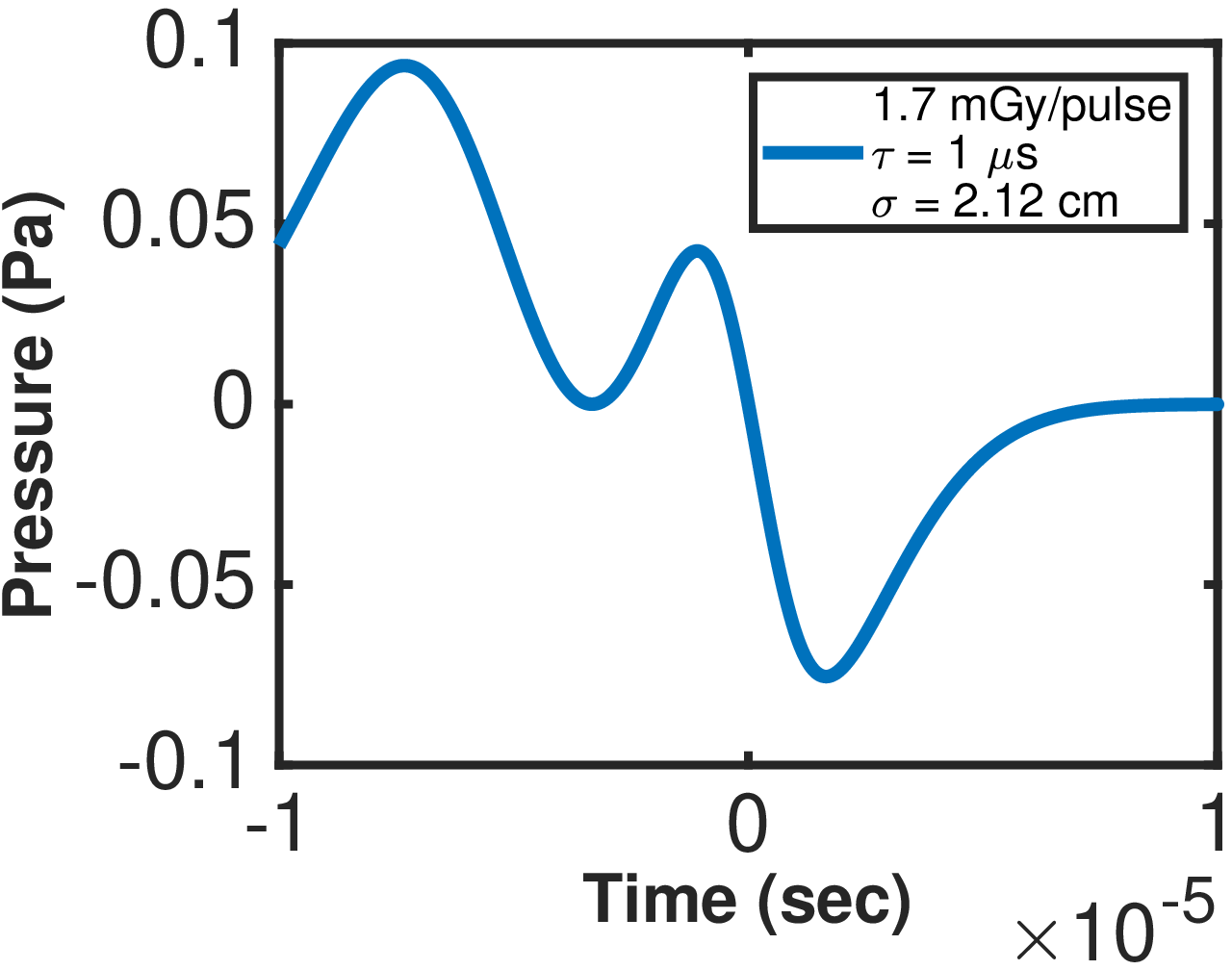}
\includegraphics[scale=0.55]{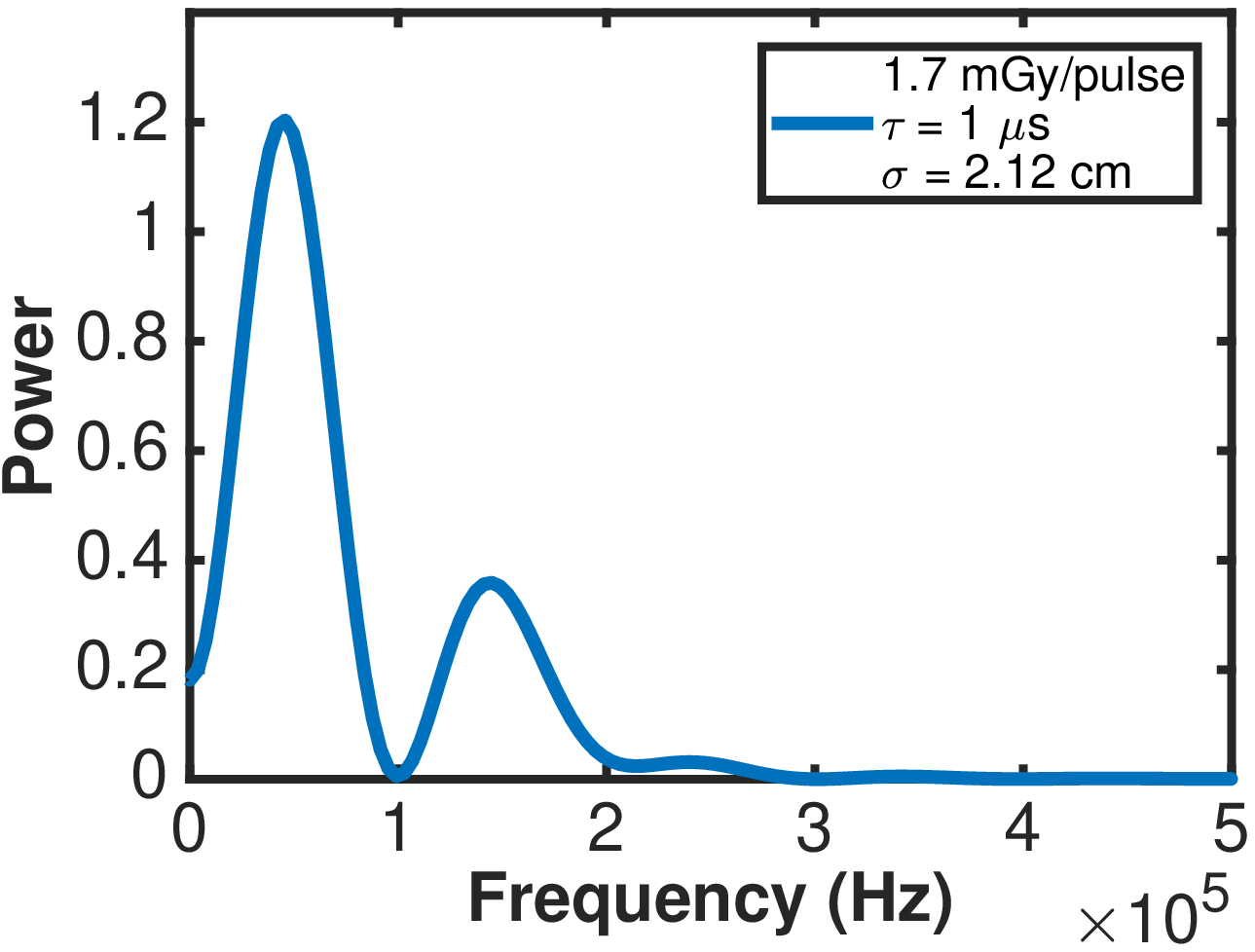}
\caption{\bf Therapeutic XA signal}
\label{fig::therapeutic_XA}
\end{figure}

\begin{figure}[!h]
\includegraphics[scale=0.55]{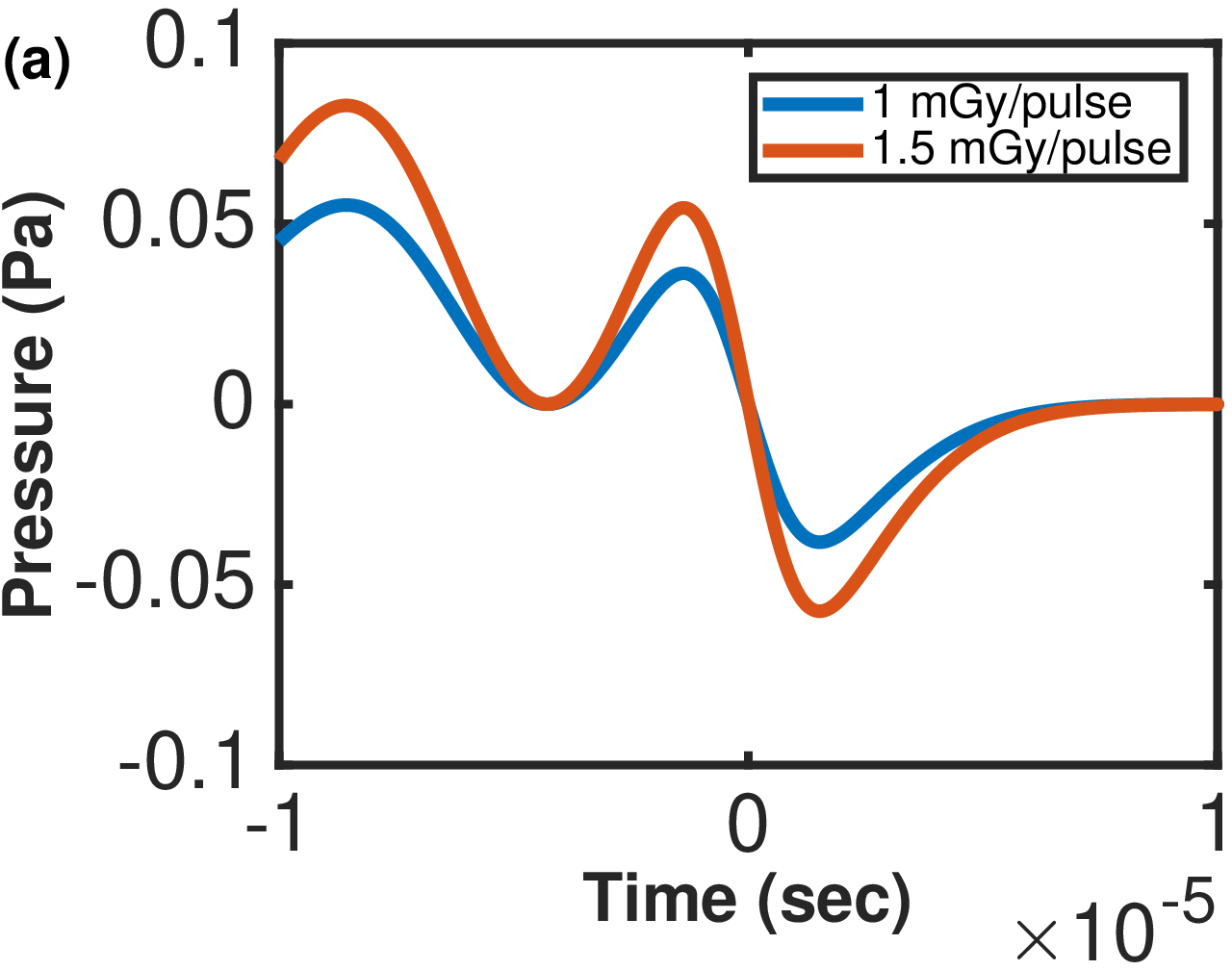}
\includegraphics[scale=0.55]{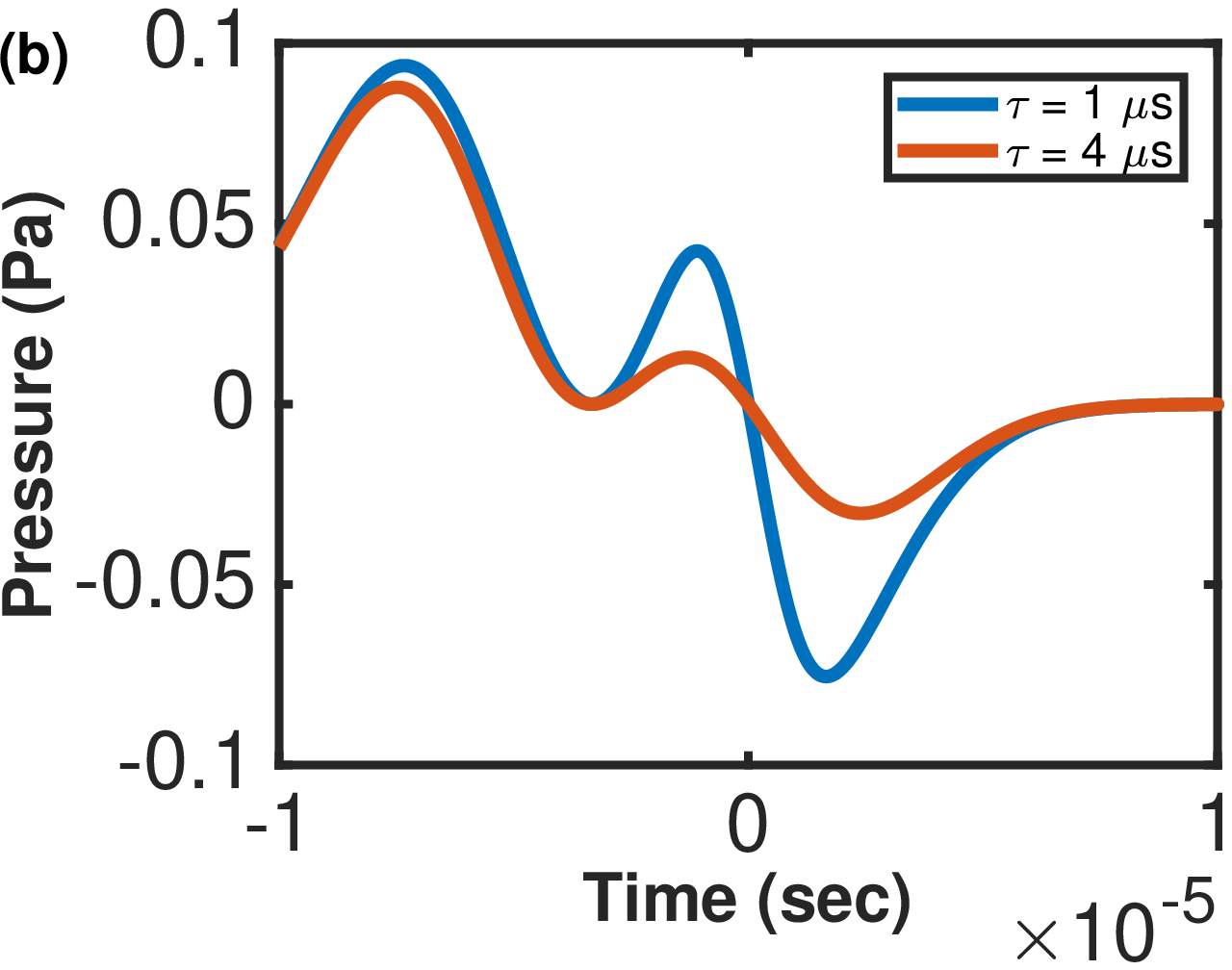}
\includegraphics[scale=0.6]{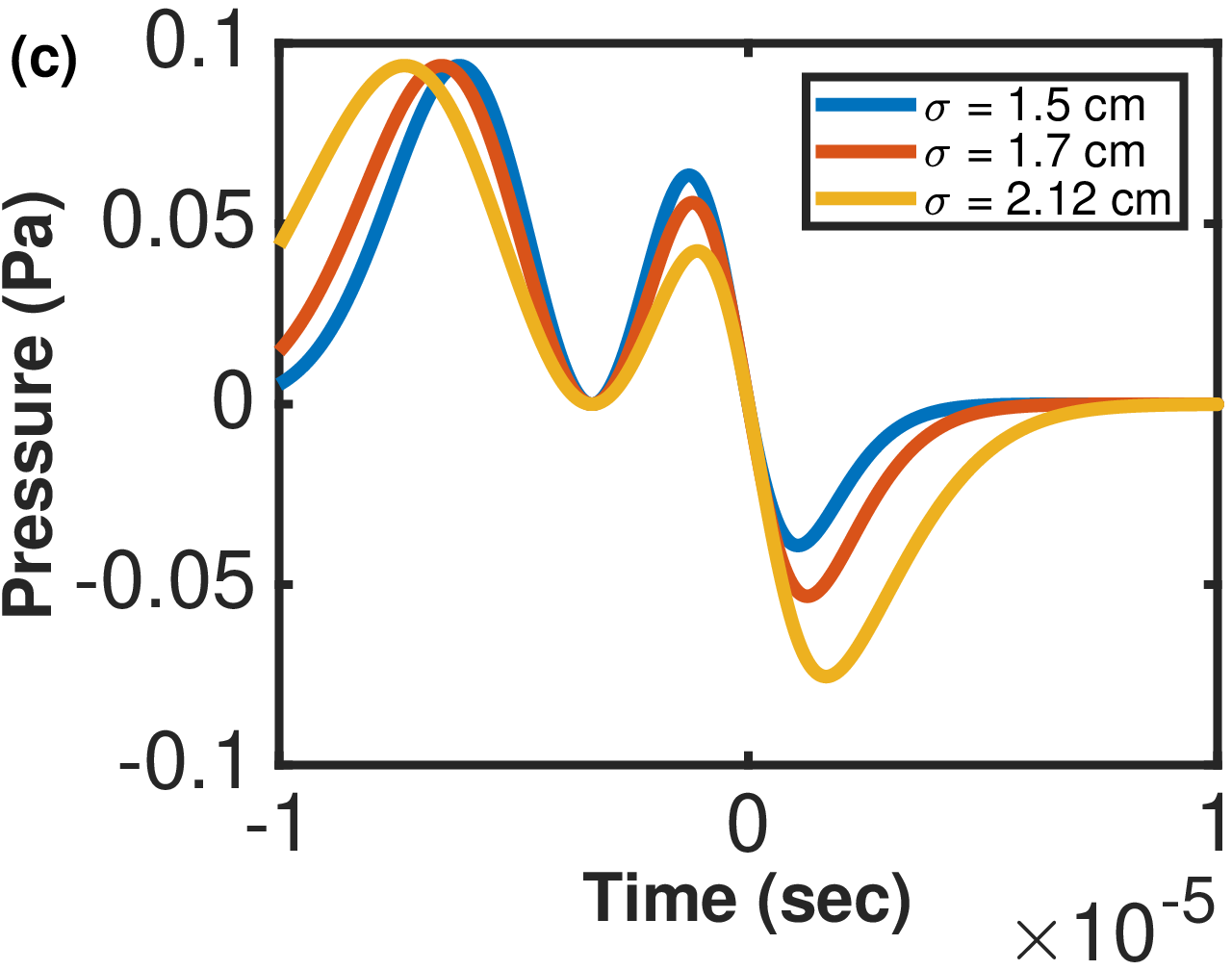} 
\caption{\bf Dependence of XA signal on Absorption Parameters}
\label{fig::pressure_dependencies}
\end{figure}

A therapeutic XA signal as being defined by the above mentioned parameters is found to have two crests and troughs corresponding to the compression and rarefaction regions separated by negative and positive dose gradients at the two boundaries of the uniform dose distribution. Pressure waveform as being observed by a transducer at a distance $\eta_{ct}=10$ cm from the center of the sphere is simulated by Eq~(\ref{pres_3d_sphere}) in time and frequency domain as it is shown in the Fig~\ref{fig::therapeutic_XA}. Based on the power spectrum of the signal which is yielded by applying FFT to the waveform in time domain and determines the central frequency as $45.8$ kHz, $NEP$ is calculated from Eq~(\ref{NEP_equation}) as $2.2$ mPa. Further, acoustic signals induced by X-ray pulses delivering 1.7 mGy dose per pulse with pulse durations of $1-3~\mu s$ at a fixed absorption cross section of $2.12~cm$ and those with a fixed pulse duration of $1~\mu s$ with different absorption cross sections defined by different width parameters of 1.5, 1.7, and 2.12 cm are obtained as given in Fig~\ref{fig::pressure_dependencies}.

\begin{figure}[!h]
\includegraphics[scale=0.55]{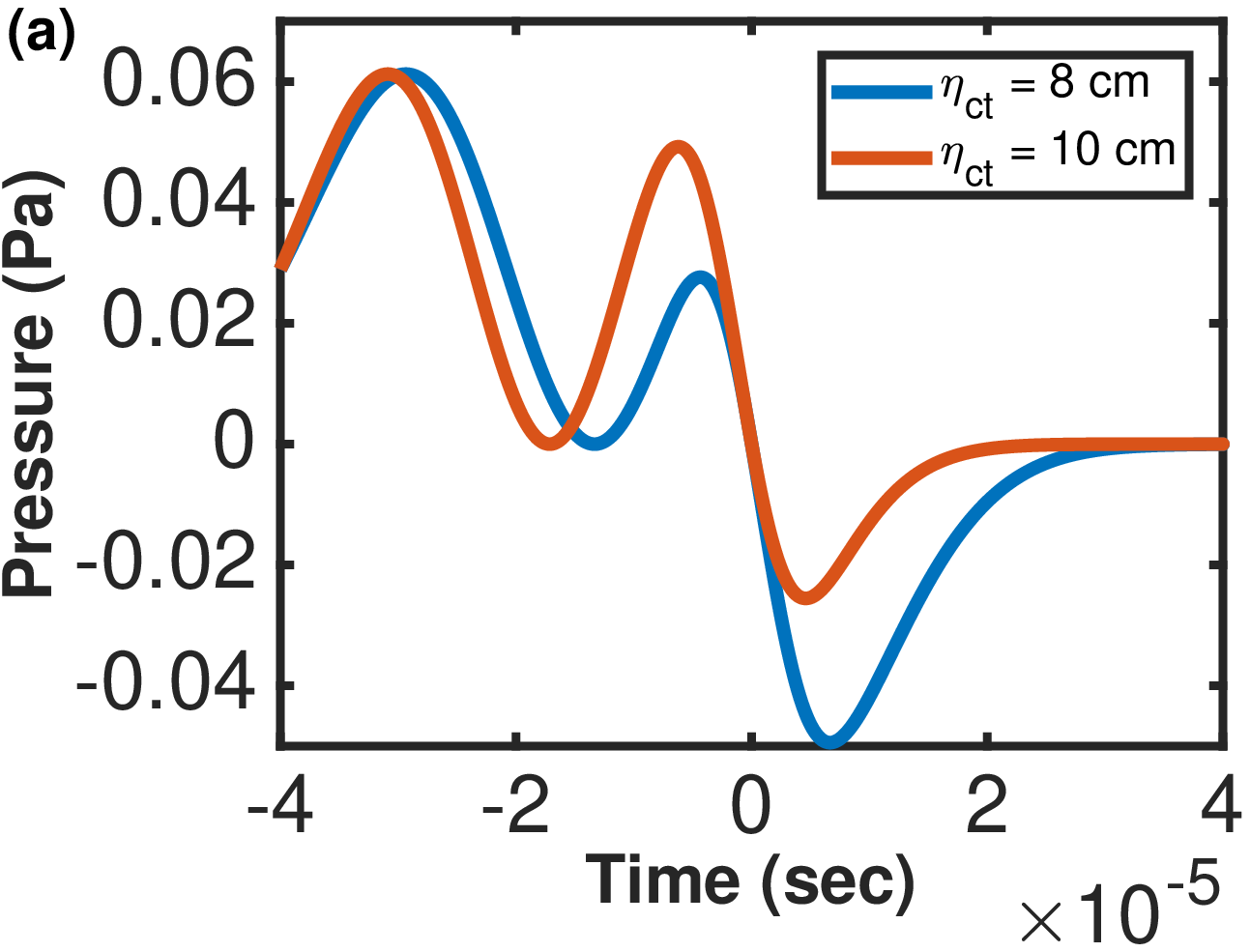}
\includegraphics[scale=0.55]{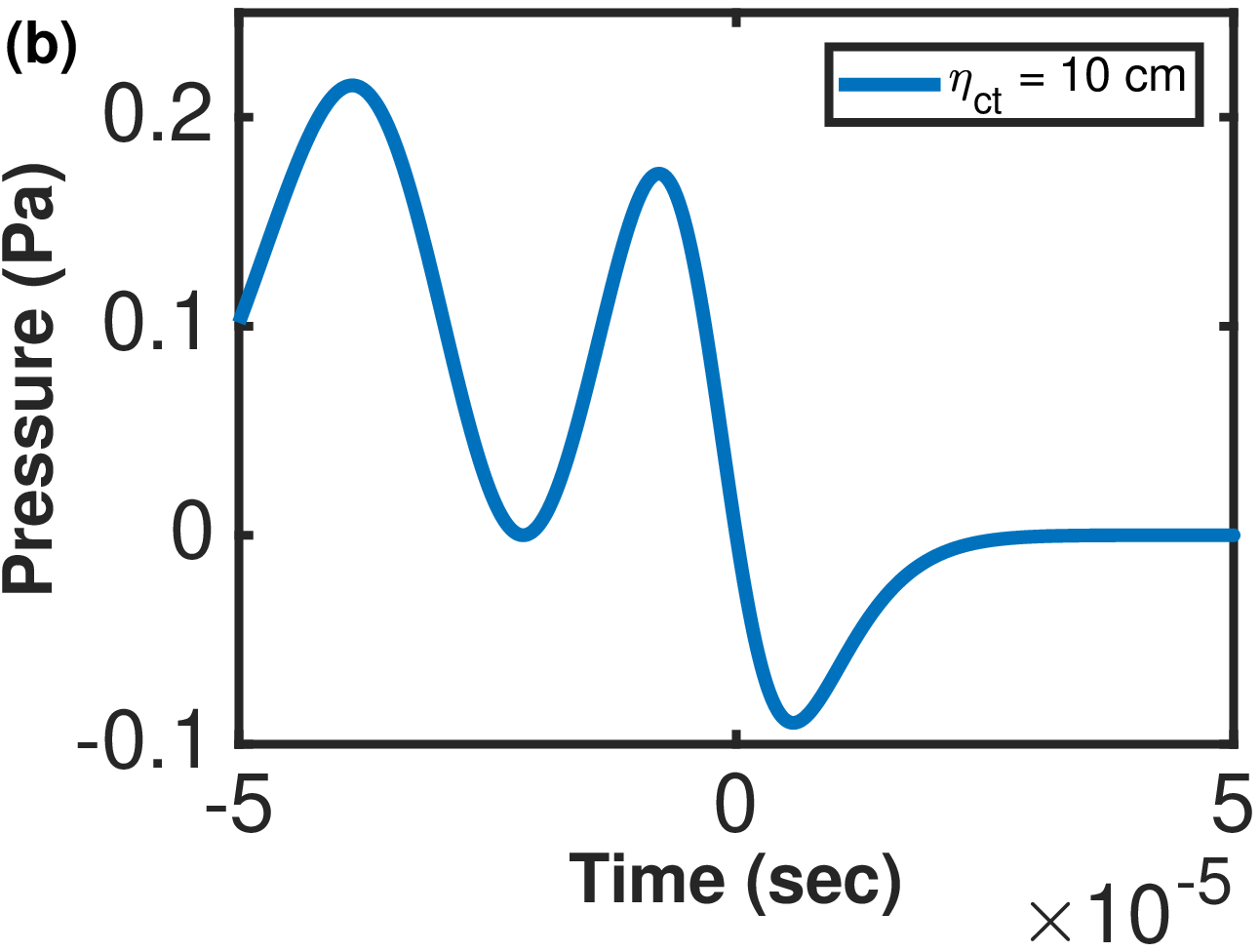}
\captionsetup{justification=raggedright,singlelinecheck=false}
\caption{\bf Simulation of Previous XA Experiments. (a) \citet{hickling2016} with transducer located at different distances of $\eta_{ct}$ from the center of the field (b) \citet{medphysxing} with irradiation realized on a pure water.}
\label{fig::previousXA}
\end{figure}

Qualification of our model is tested by simulating the XA waves obtained in empirical studies of \citet{medphysxing}, and \citet{hickling2014,hickling2016}. Necessary parameters to simulate the acoustic signals based on Eq~(\ref{pres_3d_sphere}) are given in the Table \ref{table::simulate_other_studies} where $D_{pp}$, $\tau$, and F.S. denote dose per pulse, pulse duration, and treatment field size. \citet{medphysxing} operated 10 MV linac to induce X-ray pulses from a lead rod embedded in a chicken breast tissue at a dose rate $30$ Gy/min and the pulse repetition frequency of $128$ pps from which the corresponding dose per pulse value is calculated as $3.9$ mGy. \citet{hickling2014} operated 6 MV linac at a pulse repetition of $180$ Hz to induce acoustic waves from a lead block suspended in a water tank in ``clinically relevant radiotherapy dosimetry situations" wherefore \cite{typesaccelerator} dose per pulse value can be taken as being in the range $1$ to $2$ mGy/pulse, and recently \citet{hickling2016} observed and simulated the XA signals induced from various dose distributions formed by primary collimator (or multi-leaf collimator for non-standard fields) inside pure water due to 4 $\mu s$ X-ray pulses delivering 1.11 mGy/pulse to make image reconstruction. We calculated the corresponding width parameters $\sigma$ of the Gaussian absorption profiles from the field size dimension by using Eq~(\ref{eqn::sigma}). Consequently, we have simulated the corresponding induced acoustic signals using our analytic model as shown in Fig~\ref{fig::previousXA}. Validation of our model is mainly realized by comparing XA signal in Fig~\ref{fig::previousXA}(a) with the corresponding empirical signal~\cite{hickling2016} by means of shape and relative amplitude. Other experimental XA signals~\cite{medphysxing,hickling2014} as being induced from the irradiation of the metal blocks are not suitable for a consistent comparison. However, assuming that the medium of irradiation will not significantly change the generic shape of the therapeutic XA signal since it is restrained by the medical range of the operational parameters in which those experiments are performed, qualitative comparison of the waveforms is made. Among these studies, only \citet{hickling2014} shows experimental therapeutic XA signal induced from the therapeutic X-ray irradiation of a lead block in physical units of 0.2 Pa maximum value, whereas in the other studies experimental pressure amplitudes are given in arbitrary units. This value serves as a definite upper limit for the amplitude of the XA signal induced from pure water that is simulated in this study since density of lead is ten times larger than water.

\begin{table}[!ht]
\centering
\caption{\bf Parameters to Simulate Empirical Studies}
\label{table::simulate_other_studies}
\begin{tabular}{|l|c|c|c|}
\hline
                           & \multicolumn{1}{l|}{\citet{medphysxing}} & \multicolumn{1}{l|}{\citet{hickling2014}} & \multicolumn{1}{l|}{\citet{hickling2016}} \\ \hline
\textbf{Energy (MeV)}      & 10                                          & 6                                          & 10                                             \\ \hline
\textbf{$\mathbf{D_{pp}}$ (mGy/pulse)}   & 3.9 *                                       & 1-2**                              & 1.11                                           \\ \hline
\textbf{$\mathbf{\tau}$ $\mathbf{(\mu s)}$}  & $5 $                                   & $4 $                                      & $4$                                       \\ \hline
\textbf{F.S. $\mathbf{(cm^2)}$}     & 4x4                                         & 10x10                                          & 4x4                                            \\ \hline
\textbf{$\mathbf{\sigma}$ $\mathbf{(cm)}$ ***} & 1.7                                         & 4.25                                           & 1.7                                            \\ \hline
\end{tabular}
\begin{tablenotes} 
\small
\item * calculated from Dose rate (Gy/min) and pulse repetition frequency.
\item ** in clinical range
\item *** calculated from field size by using Eq (\ref{eqn::sigma}).
\end{tablenotes}
\end{table}

\section*{Discussion and Conclusion}

Therapeutic XA signal induced from the surface of the uniform spherical dose distribution due to X-ray irradiation onto $5x5$ $cm^2$ field of the water surface by $1~\mu s$ pulses delivering $1.7$ mGy/pulse is simulated in time and frequency domain as given in Fig~\ref{fig::therapeutic_XA}, and also with different operation parameters in the medical range as given in Fig~\ref{fig::pressure_dependencies}. The linear dependence on dose rate and pressure amplitude is evident in Fig~\ref{fig::therapeutic_XA}(a) and in Fig~\ref{fig::pressure_dependencies}(a), and the x-axis limits are deliberately fixed to reveal the flattening effect of increasing pulse duration on XA signal in Fig~\ref{fig::pressure_dependencies}(b) which can be related to the decrease in the area of heat evolution due to increased rate of thermal relaxation. Lastly, assuming that the same dose is delivered by the pulses of the same length, the shape of the waveform undergoes just a phase shift and a displacement when the absorption cross section which is defined by the width parameter of the Gaussian fit to the treatment field is changed as in Fig~\ref{fig::pressure_dependencies}(c). 

To be able to justify the above observations, we simulated the proposed analytic model Eq (\ref{pres_3d_sphere}) by imposing the parameters used in previous XA studies. Sanity check of our model is realized through comparisons in waveforms, amplitudes, and power spectrum. Simulated therapeutic XA signal from the dose gradient at the treatment edges inside water is given in Fig~\ref{fig::previousXA}(a) as observed from different transducer locations, and it has the same waveform with the corresponding empirical signal~\cite{hickling2016} in which rarefaction preceding compression and compression preceding rarefaction regions are identified by crest-through and through-crest portions of the waveform which are caused by the positive and negative dose gradients at the two boundaries. Hence, we see that relative amplitude between a crest and trough gives the dose gradient information at a boundary. Experimental therapeutic XA waveforms induced from lead blocks~\cite{medphysxing,hickling2014} are also found to be the similar in shape to the analytically simulated signal. Secondly, we note that the amplitude of analytically simulated therapeutic XA signal induced from pure water has an amplitude between 50-100 mPa in the medical range of dose delivery at 1-2 mGy/pulse. Based on the previous experiments, the only conclusion we can make from this is that this amplitude range is consistent with the upper limit of 0.2 Pa set by the experimental amplitude of XA signal induced from a lead block~\cite{hickling2014}. Nevertheless, an important notice in the analytically simulated therapeutic XA signal as given in Fig~\ref{fig::therapeutic_XA} is that the relative amplitude between a crest and trough is approximately the same for two boundaries consistently with the experimental and k-wave simulated signals. Therefore, it can be further normalized based on the experimental measurements of real pressure amplitudes if necessary. Lastly, central frequency of the FFT spectrum of therapeutic XA signal is found to be $f_0=45.8$ kHz consistently with the previous results~\cite{medphysxing,hickling2014} with an error of 2.7\%. Besides, qualitative two-frequency content of the spectrum is similarly observed by~\cite{medphysxing}. The use of band-pass filter however filtered out the higher frequency in \citet{hickling2014}. An interesting but experimentally unjustified observation based on our simulations is that the waveform undergoes a phase shift and displacement as we move the transducer by preserving relative pressure amplitude as in Fig~\ref{fig::previousXA}(a) which might be accounted by the dispersive nature of sound waves. These show that analytic approach can be considered as an alternative to model therapeutic XA signals once the explicit experimental verification which may bring further corrections is also provided.

Since \citet{bowen91} observed the emission of induced acoustic signals from water due to therapeutic X-ray dose gradient at the treatment edges, succeeding XA studies \cite{hicklingfeasibility,medphysxing,hickling2014,hickling2016,screp} focused on imaging different objects inside water-based mediums. Among these studies, only \citet{medphysxing} recorded therapeutic X-ray induced acoustic signal from the tissue by operating 10 MV linac at 30 Gy/min which is a quite larger dose rate than applied in radiotherapy, and \citet{hickling2016} obtained therapeutic XA signal from the water medium which can be also considered as a tissue model, whereas other studies obtained XA signal from the irradiation of lead blocks due to its larger density causing relatively large amplitude for the acoustic signal. Physical magnitude of therapeutic XA pressure wave in tissue has still been not experimentally established although \citet{bowen91} predicted that it should be clearly visible in water at total doses less than 1 Gy. This study provides for the first time with the quantitative parametrization of radiotherapeutic XA waves. It verifies the feasibility of observing acoustic signal from radiotherapeutic photon beams based on the pressure amplitudes and NEP of an experimental transducer with an effective central frequency 50 kHz and size $3~cm^2$.

This analytic exploration of pressure-absorption dependence can be useful both for imaging and real-time dosimetry purposes, once XA signal is detected. Putting X-ray imaging into clinical practice via XA waves requires the minimization of the absorbed dose, and hence the X-ray beam energy is restrained. Having the explicit relation between pulse duration and ultrasound signal, a proper choice for the X-ray source can be made to obtain optimum signal over the noise level. On the other hand, the fact that any fractional change in absorption cross section of tissue compositions translates into change in the induced pressure profile~\cite{screp} can be used to monitor whether the irradiated region is normal or cancerous. This might enable simultaneous recognition of the amount of dose which is related with the amplitude of the XA signal in healthy tissue and hence immediate intervention to therapy session. Nevertheless, due to the uniform medium assumption, absorption cross section is solely defined by the Gaussian width parameter $\sigma$ determined by the field size on the surface by using the fact that X-ray absorption coefficient is the inverse of the minimum dimension of the area of heat evolution, in this study. This parameter can be important to distinguish acoustic signals induced in mediums with different characteristics. `Characteristics of the medium'/acoustic signal dependence is previously pointed out by \citet{screp} although it has not been proven by experiment or simulation. Depending on the medium characteristics, probability of absorption is governed by the absorption cross section which is an explicit parameter in our model. Hence, we argue that succeeding change of the pressure profile due to the change in the width parameter $\sigma$ of the Gaussian curve modeling the heating function $H(r,t)$ should mimic the medium/XA signal dependence. Once this dependence is experimentally observed, our claim, i.e. whether the dependence can be simulated by varying $\sigma$, should be tested. On the other hand, we point out that any parameter dependence other than the absorbed dose is implicit in the previous simulation model~\cite{hickling2016}, and hence established XACT simulation techniques by no chance can simulate this dependence even if it will be experimentally observed. Similar to the characterization of a bone with osteoporosis using photoacoustic waveforms~\cite{feng2015characterization}, an experiment that measures XA signals in a partially cancerous tissue might be considered for this purpose. 
 
An analytic model that can fast and simply simulate acoustic waves induced by radiotherapeutic X-rays is proposed for the first time. Considering the previous studies on this subject, we believe that the significance of this study is the foundation of a novel and self-contained analytic approach to simulate the therapeutic X-ray acoustic waves based on the physical parametrization of the energy transfer process. This not only provides a better understanding of the physical phenomena underlying the medical technique in terms of the medically relevant parameters such as field size, pulse duration, absorbed dose per pulse etc. together with the physical assumptions used to obtain a solution to the photo-acoustic equation, but also brings consistent simulation results with previous experimental and k-Wave results. This model can be easily implemented and modified for a quantitative investigation of any X-ray acoustic study, and hence it is an alternative tool to numerical approach. Further experimental research with more quantitative analysis will improve physical modeling in this area.

\section*{Disclosure of Conflicts of Interest}

The authors have no relevant conflicts of interest to disclose.

\end{document}